\documentclass[aps,twocolumn,prl,showpacs,showkeys,preprintnumbers,superscriptaddress,nobibnotes,floatfix,longbibliography,notitlepage,nofootinbib]{revtex4-2}

\usepackage{amsmath}
\usepackage{amsfonts}
\usepackage{amssymb}
\usepackage{graphicx}
\usepackage[colorlinks=true,allcolors=blue]{hyperref}
\usepackage{relsize}
\usepackage{gensymb}
\usepackage{fontawesome}
\usepackage[capitalize]{cleveref}
\usepackage[dvipsnames,table]{xcolor}

\usepackage{slashed,multirow,relsize,soul,feynmp-auto,tikz}
\usepackage{color}
\usepackage{mathrsfs} % pretty maths
\usepackage{amsmath}
\usepackage{cancel}
\usepackage{bbold}
\usepackage{mathrsfs}
\usepackage{braket}
\usepackage{physics}
\usepackage{multirow}
\usepackage{xspace}
\usepackage{etoolbox}
\usepackage{comment}
\usepackage{siunitx}
\usepackage{fontawesome} % for code link icons
\definecolor{blue-violet}{rgb}{0.33, 0.17, 0.89}

\newcommand{\journalstyle}{APS}

\begin{document}

\title{TAMBO: A Deep-Valley Neutrino Observatory}

% \usepackage{etoolbox}
% Define journal type (options: APS, AIP, IEEE, ELSEVIER, STANDARD)
%\newcommand{\journalstyle}{APS} % Change this to AIP, IEEE, ELSEVIER, STANDARD as needed

\ifdefstring{\journalstyle}{APS}{
\author{Carlos A. Argüelles}
\affiliation{Department of Physics \& Laboratory for Particle Physics and Cosmology, Harvard University, Cambridge, MA 02138, USA}

%\author{Jaime Alvarez-Mu\~niz}
%\affiliation{Instituto Galego de F'{\i}sica de Altas Enerx'{\i}as (IGFAE), Universidade de Santiago de Compostela, 15782 Santiago de Compostela, Spain}

\author{José Bazo}
\affiliation{Sección Física, Departamento de Ciencias, Pontificia Universidad Católica del Perú, Apartado 1761, Lima, Perú}

\author{Christopher Briceño}
\affiliation{Electrical and Mechatronic Department, Universidad de Ingenieria y Tecnologia, Barranco 15063, Perú}

\author{Mauricio Bustamante}
\affiliation{Niels Bohr International Academy, Niels Bohr Institute, University of Copenhagen, DK-2100 Copenhagen, Denmark}

\author{Saneli Carbajal}
\affiliation{Universidad del Pacifico, Jr. Gral, Jirón Luis Sánchez Cerro 2141 Lima, Jesús María 15072, Peru}

\author{Víctor Centa}
\affiliation{Instituto de Radioastronomía, Pontificia Universidad Católica del Perú, San Miguel, Lima 15088, Perú}

\author{Jaco de Swart}
\affiliation{Department of Physics \& Program in Science, Technology, and Society, Massachusetts Institute of Technology, Cambridge, MA 02139, USA}

\author{Diyaselis Delgado}
\affiliation{Department of Physics \& Laboratory for Particle Physics and Cosmology, Harvard University, Cambridge, MA 02138, USA}

\author{Tommaso Dorigo}
\affiliation{Lule\aa \, University of Technology, 97187 Lule\aa, Sweden}
\affiliation{INFN, Sezione di Padova, Italy}

\author{Anatoli Fedynitch}
\affiliation{Institute of Physics, Academia Sinica, Taipei City, 11529, Taiwan}

\author{Pablo Fern\'andez}
\affiliation{Donostia International Physics Center (DIPC), San Sebasti\'an/Donostia, E-20018, Spain}

\author{Alberto M. Gago}
\affiliation{Sección Física, Departamento de Ciencias, Pontificia Universidad Católica del Perú, Apartado 1761, Lima, Perú}

\author{Alfonso García}
\affiliation{Instituto de Física Corpuscular (IFIC), CSIC and Universitat de València, 46980 Paterna, València, Spain}

\author{Alessandro Giuffra}
\affiliation{Electrical and Mechatronic Department, Universidad de Ingenieria y Tecnologia, Barranco 15063, Perú}

\author{Zigfried Hampel-Arias}
\affiliation{Los Alamos National Laboratory}

\author{Ali Kheirandish}
\affiliation{Department of Physics \& Astronomy, University of Nevada, Las Vegas, NV 89154, USA}
\affiliation{Nevada Center for Astrophysics, University of Nevada, Las Vegas, NV 89154, USA}

\author{Jeffrey P. Lazar}
\affiliation{Centre for Cosmology, Particle Physics and Phenomenology -- CP3, Université catholique de Louvain, Louvain-la-Neuve, Belgium}

\author{Peter M. Lewis}
\affiliation{Department of Physics and Astronomy, University of Hawaii at Manoa, Honolulu, HI 96822}

\author{Daniel Menéndez}
\affiliation{Instituto de Radioastronomía, Pontificia Universidad Católica del Perú, San Miguel, Lima 15088, Perú}

\author{Marco Milla}
\affiliation{Instituto de Radioastronomía, Pontificia Universidad Católica del Perú, San Miguel, Lima 15088, Perú}

\author{Alberto Peláez}
\affiliation{Electrical and Mechatronic Department, Universidad de Ingenieria y Tecnologia, Barranco 15063, Perú}

\author{Andres Romero-Wolf}
\affiliation{Jet Propulsion Laboratory, California Institute of Technology, Pasadena, CA 91109, USA}

\author{Ibrahim Safa}
\affiliation{Department of Physics, Columbia University}

\author{Luciano Stucchi}
\affiliation{Universidad del Pacifico, Jr. Gral, Jirón Luis Sánchez Cerro 2141 Lima, Jesús María 15072, Peru}

\author{Jimmy Tarrillo}
\affiliation{Electrical and Mechatronic Department, Universidad de Ingenieria y Tecnologia, Barranco 15063, Perú}

\author{William G. Thompson}
\affiliation{Department of Physics \& Laboratory for Particle Physics and Cosmology, Harvard University, Cambridge, MA 02138, USA}

\author{Pietro Vischia}
\affiliation{Universidad de Oviedo and ICTEA, Calle San Francisco 3, 33007 Oviedo, Principado de Asturias, España}

\author{Aaron C. Vincent}
\affiliation{Department of Physics, Enigneering Physics and Astronomy, Queen's University, Kingston, ON, K7L 3N6, Canada}
\affiliation{Arthur B. McDonald Canadian Astroparticle Physics Research  Kingston ON K7L 3N6, Canada}
\affiliation{Perimeter Institute for Theoretical Physics, Waterloo ON N2L 2Y5, Canada}\affiliation{Department of Physics \& Laboratory for Particle Physics and Cosmology, Harvard University, Cambridge, MA 02138, USA}

\author{Pavel Zhelnin} \affiliation{Department of Physics \& Laboratory for Particle Physics and Cosmology, Harvard University, Cambridge, MA 02138, USA}

%\author{Jaime Alvarez-Mu\~niz}
%\affiliation{Instituto Galego de F'{\i}sica de Altas Enerx'{\i}as (IGFAE), Universidade de Santiago de Compostela, 15782 Santiago de Compostela, Spain}

}{}

\ifdefstring{\journalstyle}{AIP}{
% AIP format
\author{
Carlos A. Arg\"{u}elles$^1$,
Jos\'{e} Bazo$^2$,
Christopher Brice\~{n}o$^3$,
Mauricio Bustamante$^4$,
Saneli Carbajal$^5$,
V\'{i}ctor Centa$^6$,
Jaco de Swart$^7$,
Diyaselis Delgado$^1$,
Tommaso Dorigo$^{8,9}$,
Anatoli Fedynitch$^{10}$,
Pablo Fern\'andez$^{11}$,
Alberto M. Gago$^2$,
Alfonso Garc\'{i}a$^{12}$,
Alessandro Giuffra$^3$,
Zigfried Hampel-Arias$^{13}$,
Ali Kheirandish$^{14}$,
Jeffrey P. Lazar$^{15}$,
Peter M. Lewis$^{16}$,
Daniel Men\'endez$^6$,
Marco Milla$^6$,
Alberto Pel\'{a}ez$^3$,
Andres Romero-Wolf$^{17}$,
Ibrahim Safa$^{18}$,
Luciano Stucchi$^5$,
Jimmy Tarrillo$^3$,
William G. Thompson$^1$,
Pietro Vischia$^{19}$,
Aaron C. Vincent$^{20}$,
Pavel Zhelnin$^1$
}

\affil{
$^1$Department of Physics and Laboratory for Particle Physics and Cosmology, Harvard University, Cambridge, MA 02138, USA;
$^2$Secci\'on F\'isica, Departamento de Ciencias, Pontificia Universidad Cat\'olica del Per\'u, Apartado 1761, Lima, Per\'u;
$^3$Electrical and Mechatronic Department, Universidad de Ingenier\'ia y Tecnolog\'ia, Barranco 15063, Per\'u;
$^4$Niels Bohr International Academy, Niels Bohr Institute, University of Copenhagen, DK-2100 Copenhagen, Denmark;
$^5$Escuela de Ingenier\'ia, Universidad del Pac\'{\i}fico, Jir\'on Luis S\'anchez Cerro 2141 Lima, Jes\'us Mar\'ia 15072, Per\'u;
$^6$Instituto de Radioastronom\'ia, Pontificia Universidad Cat\'olica del Per\'u, San Miguel, Lima 15088, Per\'u;
$^7$Department of Physics \& Program in Science, Technology, and Society, Massachusetts Institute of Technology, Cambridge, MA 02139, USA;
$^8$Lule\aa\ University of Technology, 97187 Lule\aa, Sweden;
$^9$INFN, Sezione di Padova, Italy;
$^{10}$Institute of Physics, Academia Sinica, Taipei City, 11529, Taiwan;
$^{11}$Donostia International Physics Center DIPC, San Sebasti\'an/Donostia, E-20018, Spain;
$^{12}$Instituto de F\'isica Corpuscular (IFIC), CSIC and Universitat de Val\`encia, 46980 Paterna, Val\`encia, Spain;
$^{13}$Los Alamos National Laboratory;
$^{14}$Department of Physics \& Astronomy, University of Nevada, Las Vegas, NV 89154, USA;
$^{15}$Centre for Cosmology, Particle Physics and Phenomenology -- CP3, Universit\'e catholique de Louvain, Louvain-la-Neuve, Belgium;
$^{16}$Department of Physics and Astronomy, University of Hawaii at Manoa, Honolulu, HI 96822;
$^{17}$Jet Propulsion Laboratory, California Institute of Technology, Pasadena, CA 91109, USA;
$^{18}$Department of Physics, Columbia University;
$^{19}$Calle San Francisco 3, 33007 Oviedo, Principado de Asturias, Espa\~{n}a;
$^{20}$Department of Physics, Engineering Physics and Astronomy, Queen's University, Kingston, ON, K7L 3N6, Canada; Arthur B. McDonald Canadian Astroparticle Physics Research Institute, Kingston ON K7L 3N6, Canada; Perimeter Institute for Theoretical Physics, Waterloo ON N2L 2Y5, Canada
}
}{}

\ifdefstring{\journalstyle}{IEEE}{
    % IEEE format
    \author{Carlos A. Argüelles\thanks{Department of Physics \& Laboratory for Particle Physics and Cosmology, Harvard University, Cambridge, MA 02138, USA} \and
    José Bazo\thanks{Sección Física, Departamento de Ciencias, Pontificia Universidad Católica del Perú, Apartado 1761, Lima, Perú} \and
    Christopher Briceño\thanks{Electrical and Mechatronic Department, Universidad de Ingeniería y Tecnología, Barranco 15063, Perú} \and
    Mauricio Bustamante\thanks{Niels Bohr International Academy, Niels Bohr Institute, University of Copenhagen, DK-2100 Copenhagen, Denmark} \and
    Saneli Carbajal\thanks{Escuela de Ingeniería, Universidad del Pacífico, Jirón Luis Sánchez Cerro 2141, Jesús María 15072, Lima, Perú} \and
    Víctor Centa\thanks{Instituto de Radioastronomía, Pontificia Universidad Católica del Perú, San Miguel, Lima 15088, Perú} \and
    Jaco de Swart\thanks{Department of Physics \& Program in Science, Technology, and Society, Massachusetts Institute of Technology, Cambridge, MA 02139, USA} \and
    Diyaselis Delgado\footnotemark[1] \and
    Tommaso Dorigo\thanks{Luleå University of Technology, 97187 Luleå, Sweden} \thanks{INFN, Sezione di Padova, Italy} \and
    Anatoli Fedynitch\thanks{Institute of Physics, Academia Sinica, Taipei City, 11529, Taiwan} \and
    Pablo Fernández\thanks{Donostia International Physics Center DIPC, San Sebastián/Donostia, E-20018, Spain} \and
    Alberto M. Gago\footnotemark[2] \and
    Alfonso García\thanks{Instituto de Física Corpuscular (IFIC), CSIC and Universitat de València, 46980 Paterna, València, Spain} \and
    Alessandro Giuffra\footnotemark[3] \and
    Zigfried Hampel-Arias\thanks{Los Alamos National Laboratory} \and
    Ali Kheirandish\thanks{Department of Physics \& Astronomy, University of Nevada, Las Vegas, NV 89154, USA} \and
    Jeffrey P. Lazar\thanks{Centre for Cosmology, Particle Physics and Phenomenology -- CP3, Université catholique de Louvain, Louvain-la-Neuve, Belgium} \and
    Peter M. Lewis\thanks{Department of Physics and Astronomy, University of Hawaii at Manoa, Honolulu, HI 96822, USA} \and
    Daniel Menéndez\footnotemark[6] \and
    Marco Milla\footnotemark[6] \and
    Alberto Peláez\footnotemark[3] \and
    Andres Romero-Wolf\thanks{Jet Propulsion Laboratory, California Institute of Technology, Pasadena, CA 91109, USA} \and
    Ibrahim Safa\thanks{Department of Physics, Columbia University} \and
    Luciano Stucchi\footnotemark[5] \and
    Jimmy Tarrillo\footnotemark[3] \and
    William G. Thompson\footnotemark[1] \and
    Pietro Vischia\thanks{Calle San Francisco 3, 33007 Oviedo, Principado de Asturias, España} \and
    Aaron C. Vincent\thanks{Department of Physics, Engineering Physics and Astronomy, Queen's University, Kingston, ON, K7L 3N6, Canada} \thanks{Arthur B. McDonald Canadian Astroparticle Physics Research Institute, Kingston, ON, K7L 3N6, Canada} \thanks{Perimeter Institute for Theoretical Physics, Waterloo, ON, N2L 2Y5, Canada} \thanks{Department of Physics \& Laboratory for Particle Physics and Cosmology, Harvard University, Cambridge, MA 02138, USA} \and
    Pavel Zhelnin\footnotemark[1]}
}{}

\ifdefstring{\journalstyle}{ELSEVIER}{
% Elsevier format
% Author list for ELSEVIER using \author and \affil format

\author{Carlos A. Arg\"{u}elles}
\affil[1]{Department of Physics and Laboratory for Particle Physics and Cosmology, Harvard University, Cambridge, MA 02138, USA}

\author{Jos\'e Bazo}
\affil[2]{Secci\'on F\'isica, Departamento de Ciencias, Pontificia Universidad Cat\'olica del Per\'u, Apartado 1761, Lima, Per\'u}

\author{Christopher Brice\~no}
\affil[3]{Electrical and Mechatronic Department, Universidad de Ingenier\'ia y Tecnolog\'ia, Barranco 15063, Per\'u}

\author{Mauricio Bustamante}
\affil[4]{Niels Bohr International Academy, Niels Bohr Institute, University of Copenhagen, DK-2100 Copenhagen, Denmark}

\author{Saneli Carbajal}
\affil[5]{Escuela de Ingenier\'ia, Universidad del Pac\'ifico, Jir\'on Luis S\'anchez Cerro 2141, Jes\'us Mar\'ia 15072, Lima, Per\'u}

\author{V\'ictor Centa}
\affil[6]{Instituto de Radioastronom\'ia, Pontificia Universidad Cat\'olica del Per\'u, San Miguel, Lima 15088, Per\'u}

\author{Jaco de Swart}
\affil[7]{Department of Physics and Program in Science, Technology, and Society, Massachusetts Institute of Technology, Cambridge, MA 02139, USA}

\author{Diyaselis Delgado}
\affil[1]{}

\author{Tommaso Dorigo}
\affil[8]{Lule\aa\ University of Technology, 97187 Lule\aa, Sweden}
\affil[9]{INFN, Sezione di Padova, Italy}

\author{Anatoli Fedynitch}
\affil[10]{Institute of Physics, Academia Sinica, Taipei City, 11529, Taiwan}

\author{Pablo Fern\'andez}
\affil[11]{Donostia International Physics Center DIPC, San Sebasti\'an/Donostia, E-20018, Spain}

\author{Alberto M. Gago}
\affil[2]{}

\author{Alfonso Garc\'ia}
\affil[12]{Instituto de F\'isica Corpuscular (IFIC), CSIC and Universitat de Val\`encia, 46980 Paterna, Val\`encia, Spain}

\author{Alessandro Giuffra}
\affil[3]{}

\author{Zigfried Hampel-Arias}
\affil[13]{Los Alamos National Laboratory}

\author{Ali Kheirandish}
\affil[14]{Department of Physics and Astronomy, University of Nevada, Las Vegas, NV 89154, USA}

\author{Jeffrey P. Lazar}
\affil[15]{Centre for Cosmology, Particle Physics and Phenomenology -- CP3, Universit\'e catholique de Louvain, Louvain-la-Neuve, Belgium}

\author{Peter M. Lewis}
\affil[16]{Department of Physics and Astronomy, University of Hawaii at Manoa, Honolulu, HI 96822, USA}

\author{Daniel Men\'endez}
\affil[6]{}

\author{Marco Milla}
\affil[6]{}

\author{Alberto Pel\'aez}
\affil[3]{}

\author{Andres Romero-Wolf}
\affil[17]{Jet Propulsion Laboratory, California Institute of Technology, Pasadena, CA 91109, USA}

\author{Ibrahim Safa}
\affil[18]{Department of Physics, Columbia University, New York, NY 10027, USA}

\author{Luciano Stucchi}
\affil[5]{}

\author{Jimmy Tarrillo}
\affil[3]{}

\author{William G. Thompson}
\affil[1]{}

\author{Pietro Vischia}
\affil[19]{Universidad de Oviedo and ICTEA, Calle San Francisco 3, 33007 Oviedo, Principado de Asturias, Espa\~na}

\author{Aaron C. Vincent}
\affil[20]{Department of Physics, Engineering Physics and Astronomy, Queen's University, Kingston, ON, K7L 3N6, Canada}
\affil[21]{Arthur B. McDonald Canadian Astroparticle Physics Research Institute, Kingston, ON, K7L 3N6, Canada}
\affil[22]{Perimeter Institute for Theoretical Physics, Waterloo, ON, N2L 2Y5, Canada}
\affil[1]{}

\author{Pavel Zhelnin}
\affil[1]{}

}{}

\ifdefstring{\journalstyle}{CPS}{
\author[1]{Carlos A. Arg\"{u}elles}[]

\author[2]{Jos\'e Bazo}[]
\author[3]{Christopher Brice\~{n}o}[]
\author[4]{Mauricio Bustamante}[]
\author[5]{Saneli Carbajal}[]
\author[6]{V\'ictor Centa}[]
\author[7]{Jaco de Swart}[]
\author[1]{Diyaselis Delgado}[]
\author[8]{Tommaso Dorigo}[]
\author[22]{Tommaso Dorigo}[]
\author[9]{Anatoli Fedynitch}[]
\author[10]{Pablo Fern\'andez}[]
\author[2]{Alberto M. Gago}[]
\author[11]{Alfonso Garc\'ia}[]
\author[3]{Alessandro Giuffra}[]
\author[12]{Zigfried Hampel-Arias}[]
\author[13]{Ali Kheirandish}[]
\author[14]{Jeffrey P. Lazar}[]
\author[15]{Peter M. Lewis}[]
\author[6]{Daniel Men\'endez}[]
\author[6]{Marco Milla}[]
\author[3]{Alberto Pel\'aez}[]
\author[16]{Andres Romero-Wolf}[]
\author[17]{Ibrahim Safa}[]
\author[5]{Luciano Stucchi}[]
\author[3]{Jimmy Tarrillo}[]
\author[1]{William G. Thompson}[]
\author[18]{Pietro Vischia}[]
\author[19,20,21,1]{Aaron C. Vincent}[]
\author[1]{Pavel Zhelnin}[]

\affiliation[1]{
    organization={Department of Physics and Laboratory for Particle Physics and Cosmology, Harvard University},
    city={Cambridge},
    state={MA},
    postcode={02138},
    country={United States}
}

\affiliation[2]{
    organization={Secci\'on F\'isica, Departamento de Ciencias, Pontificia Universidad Cat\'olica del Per\'u},
    city={Lima},
    country={Per\'u}
}

\affiliation[3]{
    organization={Electrical and Mechatronic Department, Universidad de Ingenier\'ia y Tecnolog\'ia},
    city={Barranco},
    postcode={15063},
    country={Per\'u}
}

\affiliation[4]{
    organization={Niels Bohr International Academy, Niels Bohr Institute, University of Copenhagen},
    city={Copenhagen},
    postcode={DK-2100},
    country={Denmark}
}

\affiliation[5]{
    organization={Escuela de Ingenier\'ia, Universidad del Pac\'ifico},
    addressline={Jr. Gral, Jir\'on Luis S\'anchez Cerro 2141},
    city={Jes\'us Mar\'ia, Lima},
    postcode={15072},
    country={Per\'u}
}

\affiliation[6]{
    organization={Instituto de Radioastronom\'ia, Pontificia Universidad Cat\'olica del Per\'u},
    city={San Miguel, Lima},
    postcode={15088},
    country={Per\'u}
}

\affiliation[7]{
    organization={Department of Physics and Program in Science, Technology, and Society, Massachusetts Institute of Technology},
    city={Cambridge},
    state={MA},
    postcode={02139},
    country={United States}
}

\affiliation[8]{
    organization={Lule\aa\ University of Technology},
    postcode={97187},
    country={Sweden}
}

\affiliation[9]{
    organization={Institute of Physics, Academia Sinica},
    city={Taipei City},
    postcode={11529},
    country={Taiwan}
}

\affiliation[10]{
    organization={Donostia International Physics Center DIPC},
    city={San Sebasti\'an/Donostia},
    postcode={E-20018},
    country={Spain}
}

\affiliation[11]{
    organization={Instituto de F\'isica Corpuscular (IFIC), CSIC and Universitat de Val\`encia},
    postcode={46980},
    city={Paterna, Val\`encia},
    country={Spain}
}

\affiliation[12]{
    organization={Los Alamos National Laboratory},
    country={United States}
}

\affiliation[13]{
    organization={Department of Physics and Astronomy, University of Nevada},
    city={Las Vegas},
    state={NV},
    postcode={89154},
    country={United States}
}

\affiliation[14]{
    organization={Centre for Cosmology, Particle Physics and Phenomenology -- CP3, Universit\'e catholique de Louvain},
    city={Louvain-la-Neuve},
    country={Belgium}
}

\affiliation[15]{
    organization={Department of Physics and Astronomy, University of Hawaii at Manoa},
    city={Honolulu},
    state={HI},
    postcode={96822},
    country={United States}
}

\affiliation[16]{
    organization={Jet Propulsion Laboratory, California Institute of Technology},
    city={Pasadena},
    state={CA},
    postcode={91109},
    country={United States}
}

\affiliation[17]{
    organization={Department of Physics, Columbia University},
    city={New York},
    state={NY},
    country={United States}
}

\affiliation[18]{
    addressline={Calle San Francisco 3},
    postcode={33007},
    city={Oviedo, Principado de Asturias},
    country={Espa\~{n}a}
}

\affiliation[19]{
    organization={Department of Physics, Engineering Physics and Astronomy, Queen's University},
    city={Kingston},
    state={ON},
    postcode={K7L 3N6},
    country={Canada}
}

\affiliation[20]{
    organization={Arthur B. McDonald Canadian Astroparticle Physics Research Institute},
    city={Kingston},
    state={ON},
    postcode={K7L 3N6},
    country={Canada}
}

\affiliation[21]{
    organization={Perimeter Institute for Theoretical Physics},
    city={Waterloo},
    state={ON},
    postcode={N2L 2Y5},
    country={Canada}
}

\affiliation[22]{
    organization={INFN, Sezione di Padova},
    country={Italy}
}
}{}

\date{\today}

\begin{abstract}
Although the field of neutrino astronomy has blossomed in the last decade, physicists have struggled to fully map the high-energy neutrino sky. TAMBO, a mountain-based neutrino observatory, aims to solve that issue---and find clues of new physics along the way. 
\end{abstract}

\maketitle

\section{Introduction}
\label{sec:intro}

Neutrino astronomy is flourishing. Its roots run at least half a century deep, to the 1960s, when high-energy neutrinos became associated with newly discovered cosmic phenomena, such as quasars, stellar collapse, and the cosmic microwave background. 
Although plans for a deep underwater neutrino detector had already formed in the 1970s~\cite{Kotzer1976}, exploring the cosmos with neutrinos only became a reality in 2013, when the IceCube Neutrino Observatory discovered the first extra-galactic high-energy neutrinos---of petaelectrovolts (PeV)~\cite{IceCube:2013low}. The field has evolved rapidly since.

\begin{figure*}[t!]
    \centering
    \includegraphics[width=\textwidth]{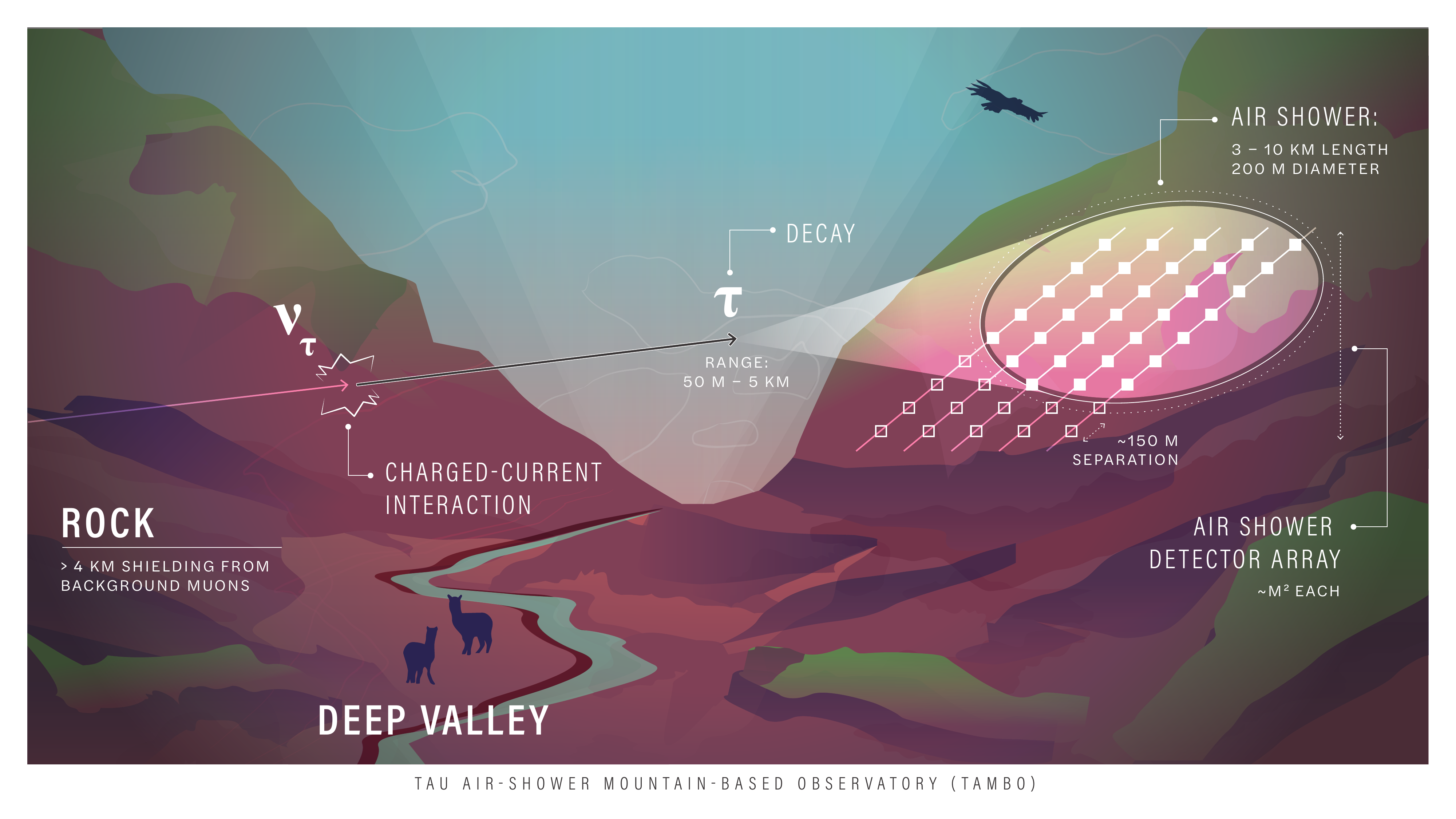}
    \caption{\textbf{\textit{Schematic of neutrino detection at TAMBO.}}
    An Earth-skimming tau neutrino undergoes a charged-current interaction within the left canyon face, producing a tau lepton.
    The tau lepton emerges from the canyon face into the valley and decays, creating an air shower.
    This air shower is detected by TAMBO, deployed on the opposite canyon face.
    }
    \label{fig:sketch}
\end{figure*}

In the last decade, the study of cosmic neutrinos has yielded significant insights into their potential sources---from the Milky Way to active galactic nuclei--and moved the field closer to finding the long-sought origin of cosmic rays~\cite{Anchordoqui:2018qom, AlvesBatista:2019tlv}.
It has furthermore enabled some of the most stringent tests of fundamental physics to date (e.g., Ref. ~\cite{IceCube:2021tdn}), along with providing hints of physics beyond the Standard Model~\cite{Carloni:2025dhv}. 

However, despite these successes, there is still a severe lack of identified cosmic high-energy neutrino sources.
Only three such sources have been identified with high significance: an active galactic nucleus, an active galaxy, and the Milky Way~\cite{IceCube:2018cha, IceCube:2022der, IceCube:2023ame}.
The neutrino flux from these sources makes up only a small fraction of the total observed cosmic neutrino flux, and the origin of most cosmic neutrinos hence remains unknown.
Additionally, the discovery of the three known neutrino sources relied on the aid of gamma-ray observations~\cite{IceCube:2018cha,IceCube:2022der}.
Neutrino sources uncorrelated with electromagnetic radiation might be abundant, yet searches for them have been unsuccessful.

 An urgent demand thus exists to develop new techniques to discover neutrino sources more effectively, in particular ones that are unbiased toward having electromagnetic counterparts. Moreover, a push towards low-cost detectors would make the field more widely accessible, support its growth, and complement the groundbreaking yet costly efforts of IceCube~\cite{IceCube:2013low}, KM3NeT~\cite{KM3NeT:2025npi}, and Baikal-GVD~\cite{Baikal-GVD:2022fis}.
Here we introduce TAMBO---the Tau Air-shower Mountain-Based Observatory---a neutrino observatory specifically designed to address these issues.

\section{Enter TAMBO}
\label{sec:detector-simulation}
   
TAMBO is designed to overcome the major challenge in identifying neutrino sources for existing experiments:  the background of atmospheric neutrinos that is roughly a million times larger than cosmic neutrino signals. 
Unlike traditional neutrino telescopes, TAMBO is developed to be sensitive to higher-energy, supra-PeV \textit{tau} neutrinos. 
At these high energies, the cosmic neutrino signal dominates over the atmospheric neutrino background, particularly for tau  neutrinos. 
As a result, any neutrino detected by TAMBO is likely to be of cosmic origin.

The key to TAMBO's concept is the use of the topography of a \textit{deep valley }(\Cref{fig:sketch}). Upon reaching Earth, high-energy cosmic tau neutrinos ($\nu_\tau$) can travel for up to hundreds of kilometers through the Earth's crust.
Each $\nu_\tau$ has a chance to interact with the rock in the crust and produce a short-lived tau lepton, the probability of which increases with neutrino energy. If this occurs in the vicinity of a deep valley, the tau lepton exits into the air, decays, and initiates an extensive air shower of particles that cross the expanse of the valley.
TAMBO aims to observe these air showers by placing an array of particle detectors on one face of the valley, allowing it to monitor the opposite face and thus detect \textit{Earth-skimming} tau neutrinos that reach it from nearly horizontal directions---realizing an idea first introduced in 1999~\cite{Fargion:1999se}.

\begin{figure}
    \centering
    \includegraphics[width=\linewidth]{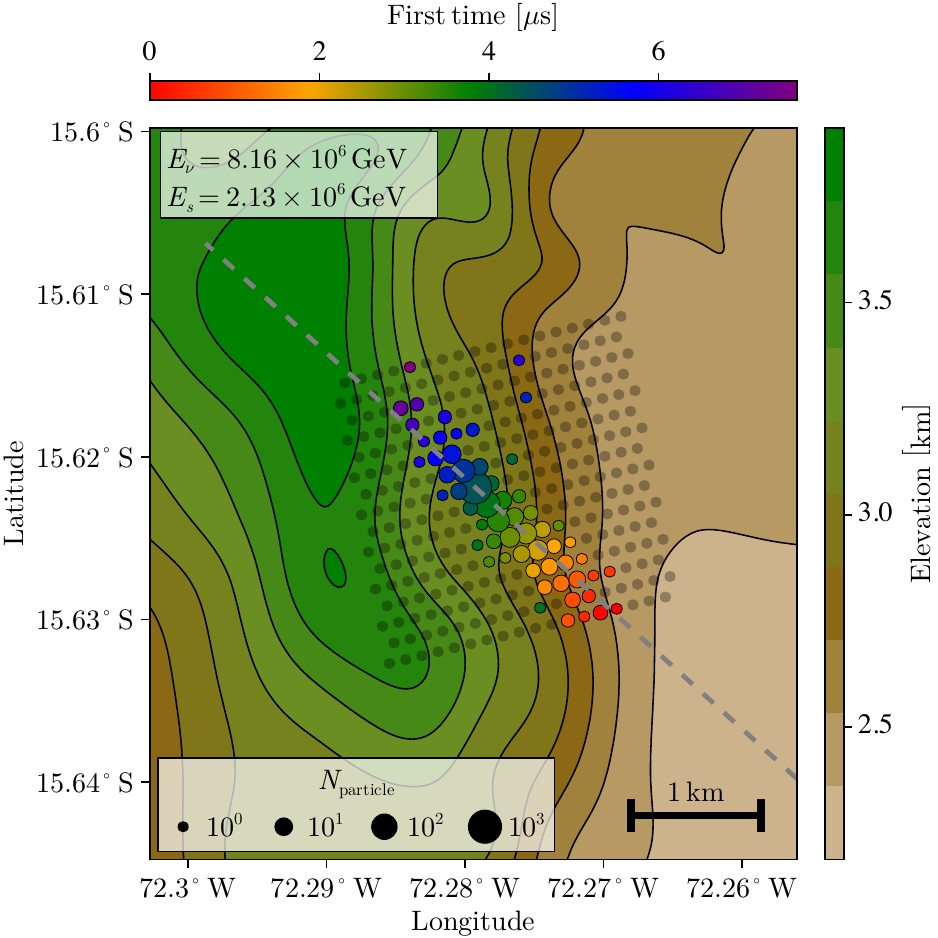}
    \caption{\textbf{\textit{Simulation of neutrino detection in TAMBO.}}
    Each detection unit in the TAMBO array is marked as a dot. A tau neutrino, with energy $E_\nu$, interacts inside the valley face near the bottom-right corner of the figure and produces a tau that, upon decaying, initiates a particle shower with energy $E_s$. 
    Detection units triggered by shower particles are displayed as colored circles, their size proportional to the number of detected particles, $N_{\rm particle}$, and their color indicating the time when the first particle hits the detector. 
    The simulated array is located in Peru's Colca Canyon, a candidate site, with an approximate depth of 1.5~km and a median distance between valley sides of 4.5~km. 
    }
    \label{fig:simulation}
\end{figure}

TAMBO is set up to determine the direction of tau neutrinos with sub-degree angular resolution, which is a crucial requisite to search for neutrino point sources. 
In its nominal configuration, TAMBO comprises 5,000~detection units spaced \SI{150}{m} apart on a triangular grid, to account for the footprint of a typical particle shower in a deep valley  (i.e., hundreds of meters).
From the number of particles incident on each detection unit and the times at which different units are hit, TAMBO infers the neutrino energy and direction.
\Cref{fig:simulation} illustrates this setup and shows the simulated detection of a neutrino with about $\SI{8}{PeV}$ of energy. 

\begin{figure}[t!]
    \centering
    \includegraphics[width=\linewidth]{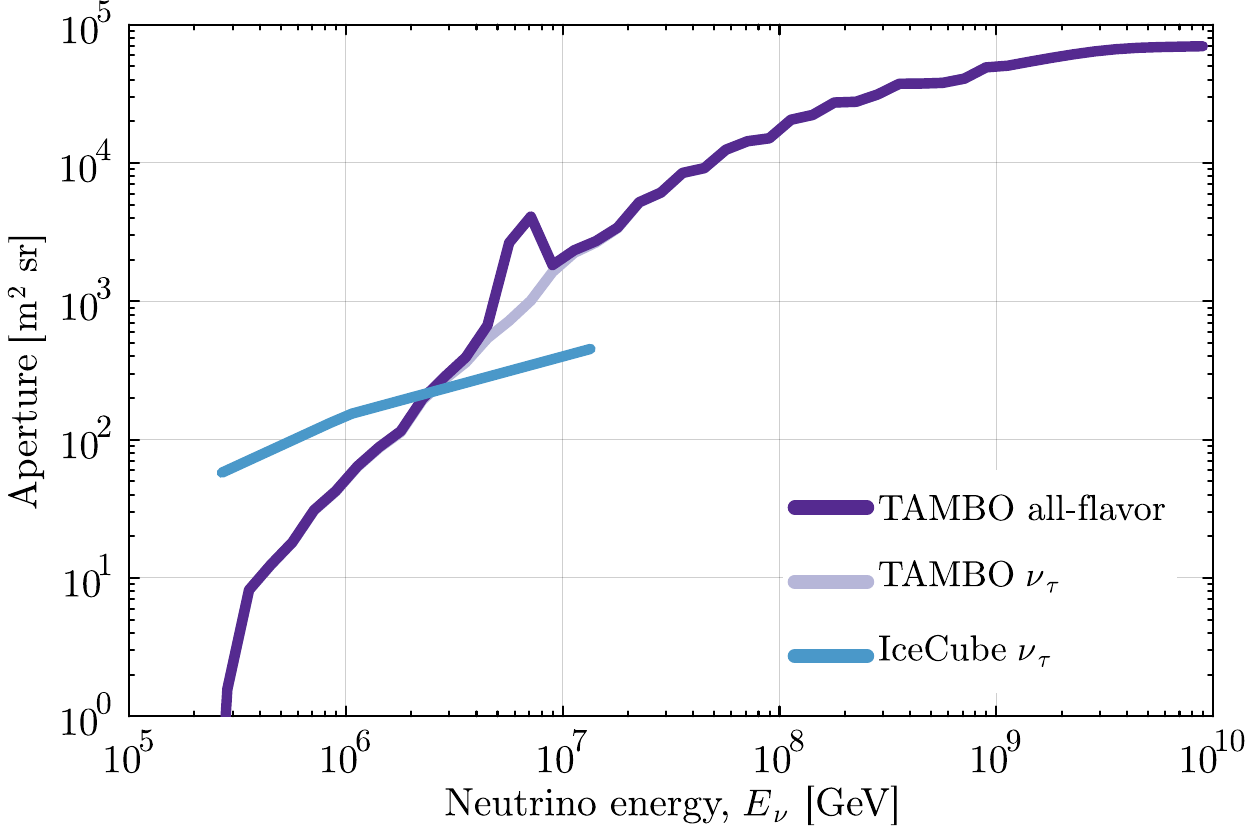}
    \caption{\textbf{\textit{Expected acceptance of TAMBO.}} 
    The acceptance is estimated by simulating an array of 5000~detection units spaced 150~m apart at a site in the Colca Canyon. The all-flavor aperture comprises contributions from neutrino types that can create a tau decay in the valley, $\nu_\tau$, $\bar{\nu}_\tau$, and $\bar{\nu}_e$ via the Glashow resonance. The contribution from $\nu_\tau$ and $\bar{\nu}_\tau$ only is also presented; the corresponding IceCube acceptance is shown for comparison.
    \label{fig:performance}}
\end{figure}

\Cref{fig:performance} shows TAMBO's aperture, which is proportional to its sensitivity to the incoming neutrino flux.
Throughout most of the energy range, the signal primarily originates from $\nu_\tau$-initiated showers. 
Around \SI{6.3}{PeV}, the signal is dominated by the Glashow resonance of $\bar{\nu}_e$ interacting with electrons, which create $W^-$ bosons that decay to tau leptons.
Above \SI{2}{PeV}, the TAMBO aperture exceeds the IceCube $\nu_\tau$ aperture, which is limited by the requirement that showers be contained within the instrumented volume.
Above $\SI{1}{EeV}$, the TAMBO aperture flattens because the boosted tau lifetime prevents it from decaying inside the valley.

In its nominal configuration, we expect TAMBO to observe, on average, 6 extragalactic neutrinos every ten years, assuming only the neutrino flux measured by IceCube in Ref.~\cite{Naab:2023xcz}, and more than 20 in ten years in more optimistic scenarios~\cite{Rodrigues:2020pli}.
While the overall number of detected neutrinos may be small compared to observatories such as IceCube, the cosmic purity of TAMBO's sample will be substantially higher: every neutrino detected by TAMBO will mark the observation of either a cosmic neutrino source or the discovery of the long-theorized cosmogenic neutrino flux.
Beyond its own science goals (see below), TAMBO thus acts as a \textit{viewfinder} for other neutrino telescopes worldwide: providing precise locations of likely neutrino sources to inspect more closely and overcoming the statistical penalties that weaken current full-sky and catalog source searches.

TAMBO furthermore benefits from a relatively low cost compared with traditional water- or ice-Cherenkov neutrino telescopes: its detection units are significantly cheaper and easier to deploy, as they do not require deep drilling in ice or deployment in the sea.
We expect TAMBO's final price tag to be around 10\% of that of IceCube. 
The performance of TAMBO scales with the number of detection units deployed, and partial goals are achievable even with a fraction of the array simulated here, making TAMBO resilient should funding become limited.
Considering how strongly TAMBO will boost the capabilities of existing neutrino telescopes to study neutrino sources, we believe it could be their most cost-effective enhancement.

A key design challenge for TAMBO's detection concept is selecting an optimal geographic site that provides the required surface area and valley width.
Several potential locations for TAMBO are currently under consideration for their technical, societal, and environmental suitability, with one strong possibility being the Colca Valley in the Peruvian Andes.
Because mountain ecosystems are generally fragile and sites can have potential histories of foreign exploitation, our collaboration is spearheading a socially and environmentally responsible approach to the site selection procedure, inspired by similar discussions taking place in other areas of astronomy~\cite{DeSwart2024}.

\section{PeV Neutrino Astrophysics}
\label{sec:AstroFront}

\textit{Discovering new astrophysical neutrino sources---}Below $\sim \SI{100}{PeV}$, neutrinos observed by TAMBO will likely originate from inside cosmic accelerators, thus pointing to the locations of potential neutrino sources.
Following an observation by TAMBO, the potential source location can be promptly investigated by the existing larger-statistics, lower-purity neutrino telescopes (IceCube, KM3NeT, BaikalGVD) using their full archival datasets. This will substantially increase the sensitivity of these telescopes by reducing the size of the trials factor, compared to contemporary source-search techniques.
With the lower threshold of these experiments, and given that the neutrino flux typically falls with neutrino energy, they will be capable of seeing many events from the TAMBO-discovered candidate source positions.
This could lead to the first neutrino-only source discovery in astronomy, without relying on observations using light or light-based catalogs.

\textit{Searching for the diffuse cosmogenic neutrino flux---}Above $\sim \SI{100}{PeV}$, any neutrinos observed by TAMBO are more likely to have originated from a diffuse flux of \textit{cosmogenic neutrinos} rather than from cosmic accelerators; these neutrinos are produced by ultra-high-energy protons scattering off the cosmic microwave background~\cite{Berezinsky:1969erk}. 
As such, TAMBO will provide crucial insights on models of cosmogenic neutrinos, helping to explain the dearth of cosmic rays with energies above the Greisen-Zatsepin-Kuzmin limit, \SI{5e19}{eV}~\cite{Greisen:1966jv,Zatsepin:1966jv}.
Our forecasts show that, after seven years of operation, TAMBO will be sensitive to optimistic cosmogenic neutrino flux models, such as that proposed by Bergman \& van Vliet~\cite{Anker:2020lre}.
This sensitivity will also enable crucial new insight into KM3NeT's recent \SI{220}{PeV} neutrino observation.
The origin of this neutrino is currently unknown~\cite{KM3NeT:2025vut}, and combined analyses with IceCube data have so far been inconclusive~\cite{KM3NeT:2025ccp}.

In addition to searching for the cosmogenic neutrino flux, TAMBO will expand upon IceCube's measurements~\cite{IceCube:2025ezc} of the diffuse astrophysical neutrino flux, thought to originate from unresolved point sources. TAMBO extends these measurements above several \si{PeV}: between \SI{1}{PeV} and \SI{1}{EeV}, its projected sensitivity to this diffuse astrophysical flux exceeds that of all present-day observatories.

\section{Messengers of Uncharted Physics}
\label{sec:NewPhysics}

\textit{Searching for new particle phenomena---}In just over a decade, neutrino astrophysics has provided a wealth of tests of the Standard Model.
Such tests include measuring the neutrino-nucleon cross section~\cite{IceCube:2020rnc}, probing neutrino interactions with dark matter~\cite{IceCube:2022clp}, searching for sterile neutrinos~\cite{IceCubeCollaboration:2024nle}, scrutinizing different neutrino mass mechanisms~\cite{Carloni:2022cqz}, testing fundamental symmetries~\cite{IceCube:2017qyp}, and setting limits on quantum gravity~\cite{ICECUBE:2023gdv}.
Despite this broad program, the strength of such tests degrades at $\gtrsim\si{PeV}$ energies due to the dearth of neutrinos detected at these energies.

TAMBO will extend these tests to energies above a few \si{PeV} while also enabling new tests of physics beyond the Standard Model by taking advantage of its $\nu_\tau$ exclusivity.
The $\nu_\tau$ is the least-observed Standard Model particle, and thus there are considerable opportunities to discover or constrain new physics associated with it.
Of particular interest is TAMBO's ability to further constrain the cosmic neutrino flux's flavor composition by combining its data with that of IceCube.
Because the flavor composition is a particularly versatile probe of new physics~\cite{Arguelles:2022tki}, such a measurement would inform multiple new-physics models.

\textit{Understanding hadronic models and cosmic-ray physics---}Because of its location in a valley, TAMBO will also uniquely contribute to understanding cosmic rays and their associated air showers, complementing existing experiments such as IceTop~\cite{IceCube:2012nn}  and the Pierre Auger Observatory~\cite{PierreAuger:2015eyc}. 

In particular, TAMBO has significant potential to address the \textit{cosmic-ray muon puzzle}: the persistent underprediction of the average number of muons by air-shower simulations, despite correctly predicting other shower properties ~\cite{Albrecht:2021cxw}. Leveraging its high-altitude location, inclined geometry, and natural geological features, TAMBO could utilize surrounding rock formations to selectively shield the electromagnetic component of air showers. This approach would enable simultaneous measurements of electromagnetic and muonic components, offering a new perspective on air shower physics at energies where discrepancies in muon predictions become evident. Resolving the muon puzzle will directly impact neutrino astronomy: the same hadronic models thought to give rise to the muon puzzle are also used to predict part of the atmospheric neutrino flux, a key background in traditional neutrino observatories.

TAMBO's sensitivity to neutrino messengers in the \si{PeV} to \si{EeV} energy range will help map the neutrino sky, search for of new physics, and unlock the door to a new era of neutrino astronomy---moving the field from deep ice and deep waters into the deep valley.
In Quechua, the indigenous language spoken in the Andean mountains, \textit{tambo} means ``inn'': it was the name of a building in the Inca empire used as a resting place for messengers traversing the mountains.
The TAMBO experiment echoes this; it will be a home for cosmic neutrino messengers that bring news from the highest-energy phenomena in the Universe.

\section{Acknowledgments}
\begin{acknowledgments}
We would like to thank Jaime Álvarez Muñiz and Enrique Zas for their initial contributions to TAMBO.
We would also like to thank the Milton Family Fund at Harvard, the Harvard Faculty of Arts and Sciences Dean's Fund for Promising Scholarship, and the Harvard-UTEC fund.
The initial development of the social aspects of a large neutrino telescope in the Peruvian Andes was supported by the Radcliffe Institute for Advanced Study at Harvard University.
CAD and WT were partially supported by the Canadian Institute for Advanced Research (CIFAR) Azrieli Global Scholars program through this work.
CAA is supported by the Faculty of Arts and Sciences of Harvard University, the National Science Foundation (NSF), the NSF AI Institute for Artificial Intelligence and Fundamental Interactions, the Research Corporation for Science Advancement, and the David \& Lucile Packard Foundation.
CAA and JL were supported by the Alfred P. Sloan Foundation for part of this work.
PV is supported by the ``Ramón y Cajal” program under Project No. RYC2021-033305-I funded by the MCIN MCIN/AEI/10.13039/501100011033 and by the European Union NextGenerationEU/PRTR.
MB is supported by \textsc{Villum Fonden} under project no.~29388.
AG is supported by the CDEIGENT grant No. CIDEIG/2023/20.
JL is a postdoctoral researcher at the Fonds de la Recherche Scientifique - FNRS.
JB and AG acknowledge the Dirección de Fomento de la Investigación (DFI-PUCP) for funding under grant CAP-PI1144.
JdS is supported by the American Institute of Physics Robert H.G. Helleman Memorial Postdoctoral Fellowship.
\end{acknowledgments}

%%%%%%%%%%%%%%%%%%%%%%%%%%%%%%%%%%%%%%%%%
\bibliographystyle{apsrev4-2}
\bibliography{main}{}
%%%%%%%%%%%%%%%%%%%%%%%%%%%%%%%%%%%%%%%%%

\end{document}